*"Quest for the non-perturbative magnetic field effects in 1000-Tesla"*

# Helical Edge Transport in the $\nu = 0$ Quantum Hall Ferromagnetic State of an Organic Dirac Fermion System

Toshihito Osada[1,2]*, Mitsuyuki Sato[2], Takako Konoike[2,3], and Woun Kang[2,4]

[1]*Department of Applied Physics, University of Tokyo, Tokyo 113-8656, Japan.*
[2]*Institute for Solid State Physics, University of Tokyo, Kashiwa, 277-8581, Japan.*
[3]*National Institute for Materials Science, Tsukuba, Ibaraki 305-0003, Japan.*
[4]*Department of Physics, Ewha Womans University, Seoul 03760, Korea.*

*E-mail: osada@issp.u-tokyo.ac.jp

**Abstract.** We experimentally confirm the $\nu = 0$ quantum Hall ferromagnetic (QHF) state, accompanied by helical edge states, in the layered organic Dirac-fermion system $\alpha$-$(ET)_2I_3$ by demonstrating helical edge transport in multilayers. The saturation of interlayer magnetoresistance (MR) in the high-magnetic-field quantum limit does not scale with the sample cross-sectional area and appears when the magnetic field is oriented along the side surface. The in-plane MR exhibits a similar angle dependence, whereas this feature disappears in the Corbino geometry where no edge channels are present. These results are consistent with helical edge transport in the multilayer QHF state. They also rule out the possibility that the observed angle-dependent MR arises from the chiral magnetic effect expected for a 3D Dirac or Weyl semimetal.

## 1. Introduction: edge (surface) states in quantum Hall ferromagnet

A layered organic conductor, $\alpha$-$(ET)_2I_3$, is known to host a two-dimensional (2D) massless Dirac-fermion system under high pressure (>1.5 GPa) [1]. The Dirac-specific $N = 0$ Landau level (LL), which carries both spin and valley degeneracy, emerges at the Dirac point. It is half-filled because the Fermi level lies exactly at the Dirac point at zero magnetic field. When the Zeeman spin splitting becomes sufficiently large compared with the LL spacing at high magnetic fields, the Fermi level enters the mobility gap between the spin-split $N = 0$ LLs, resulting in the formation of a $\nu = 0$ quantum Hall ferromagnetic (QHF) state [2]. In this $\nu = 0$ QHF phase, helical edge states with opposite spin and chirality appear along the sample edges (side surfaces) and dominate the transport phenomena, as schematically illustrated in Fig. 1.

The QHF state was first discussed in graphene, but its experimental realization has not yet been conclusively demonstrated because other quantum Hall (QH) states compete in graphene [3], where the spin splitting is relatively small. Thus, $\alpha$-$(ET)_2I_3$ is one of the most promising candidates for realizing the QHF state; moreover, unlike graphene, it naturally forms a multilayer QHF system.

In our previous theoretical work [4], we interpreted the saturation of interlayer

*"Quest for the non-perturbative magnetic field effects in 1000-Tesla"*

magnetoresistance (MR) observed at high magnetic fields in $\alpha$-(ET)$_2$I$_3$ as a manifestation of side-surface transport mediated by QHF helical edge states. We further pointed out that the probability of interlayer tunneling between edge states on adjacent layers, which gives rise to interlayer surface transport, depends on the magnetic-field orientation, exhibiting a resonant enhancement when the field is oriented parallel to the side surface. Based on these theoretical predictions, in this paper we provide experimental evidence for transport via the helical surface (edge) states of the QHF phase in $\alpha$-(ET)$_2$I$_3$.

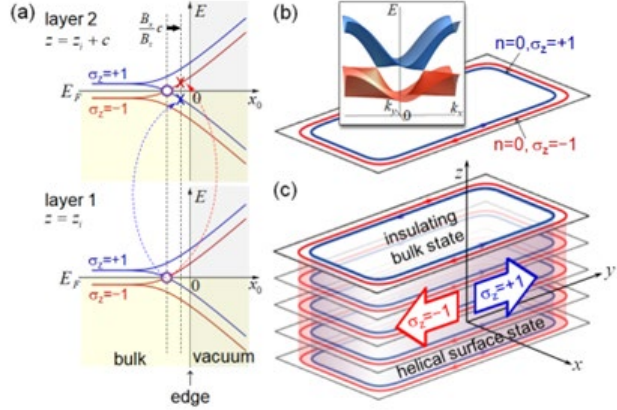

**Fig. 1.** (a) Interlayer tunneling between helical edge state on adjacent layers.
(b)(c) Helical edge state in the 2D QHF and the multilayer QHF [6].

## 2. Results: experimental evidence for edge (surface) transport

To confirm the QHF state at the high-magnetic-field quantum limit of $\alpha$-(ET)$_2$I$_3$, we performed several experiments focusing on surface transport specific to the QHF phase.

**(1) Scaling of saturating component of interlayer MR**

In $\alpha$-(ET)$_2$I$_3$ at the quantum limit, the interlayer magnetoresistance (MR) first exhibits an anomalous decrease (negative MR), followed by an increase. These behaviors can be attributed to the increase in degeneracy and the Zeeman spin splitting of the $N = 0$ LL, respectively. At higher magnetic fields, the MR increases exponentially and eventually reaches a saturation [4]. In our previous work [5], we argued that this saturation originates from additional surface transport on the side surfaces in the QHF state.

This scenario can be experimentally tested by examining whether the interlayer MR scales with the sample cross-sectional area or with the sample perimeter. To this end, we prepared two samples cut from the same large plate-like crystal so that they have identical thickness but different cross-sectional areas and perimeters. Figure 2 shows the temperature and magnetic-field dependence of the interlayer MR for the two samples, where the MR is normalized by the cross-sectional area [6].

At low temperatures and high magnetic fields, the MR clearly shows saturation. Below the saturation regime, the MR curves for the two samples nearly coincide, implying that the bulk transport scales with the sample cross-sectional area. In contrast, the MR curves deviate from each other in the saturating region. This deviation indicates the emergence of an additional metallic transport channel that does not scale with the cross section, strongly suggesting interlayer conduction mediated by metallic helical edge states in the multilayer QHF state.

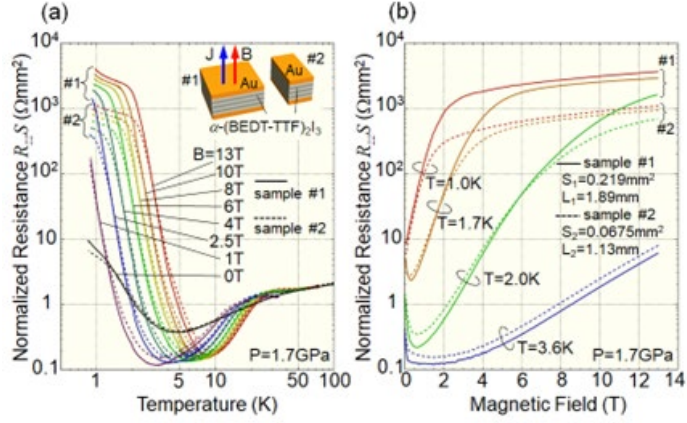

**Fig. 2.** (a) Temperature dependence and (b) magnetic field dependence of interlayer MR normalized by cross-sectional area of two samples [6].

**(2) Angle-dependence of interlayer MR**

We examine how the saturating part of the interlayer MR depends on the magnetic-field orientation. Figure 3 shows the measured interlayer MR as a function of both field strength and orientation: the distance and direction from the origin represent the field magnitude and orientation, respectively, and the color scale indicates the MR value [6]. The $z$-axis is taken to be normal to the 2D layers, while the $x$-axis is chosen arbitrarily within the plane.

Although the MR generally increases monotonically with increasing magnetic-field strength for most field orientations, it exhibits a clear saturation around the direction indicated by the dashed line. This feature is consistent with theoretical simulations of surface transport in the QHF state, which predict that the interlayer MR saturates when the magnetic field is oriented parallel to the side surface [5]. This behavior originates from resonant tunneling between helical edge states on adjacent layers when the field is aligned with the side-surface direction.

Therefore, the observed angle-dependent saturation of the interlayer MR is well explained by interlayer transport mediated by helical edge states in the multilayer QHF state.

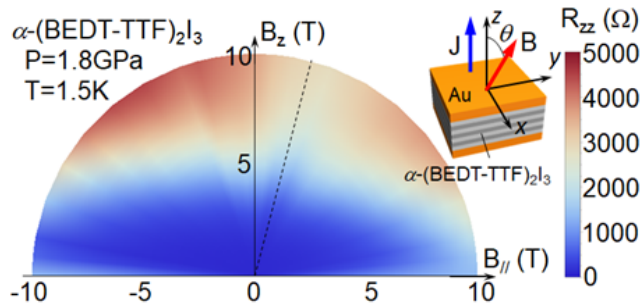

**Fig. 3.** Dependence of the interlayer MR on field strength and orientation [6].

**(3) High-magnetic-field interlayer MR**
To examine the stability of the QHF state, we measured the interlayer MR in extended magnetic fields up to 31 T using a water-cooled static magnet at the National High Magnetic Field Laboratory (NHMFL) in Tallahassee, USA. As shown in Fig. 4, the saturation behavior persists up to 31 T, at which the Zeeman energy $\mu_B B$ reaches approximately 20 K. According to theoretical studies on the stability of competing $\nu = 0$ QH states in graphene, the QHF state is expected to be the ground state in the high-field limit because its stability increases with increasing Zeeman energy [3]. Therefore, the persistence of the saturation up to such high magnetic fields strongly suggests that it originates from the QHF state.

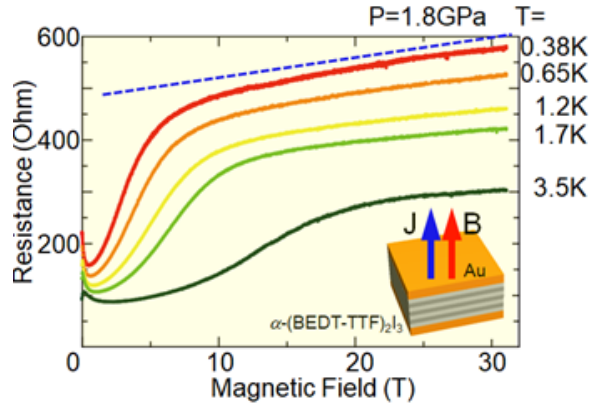

**Fig. 4.** Interlayer MR at higher magnetic fields..

**(4) In-plane edge transport**
We also investigated the in-plane transport as a function of magnetic-field orientation. For this purpose, standard and Corbino-type electrode configurations were fabricated on two pieces cut from the same plate-like crystal. In the standard configuration, the magnetic field was rotated within the plane perpendicular to the current direction (*x*-direction), and its orientation was measured by the angle θ from the normal (*z*-direction) of the 2D layers.

Figures 5(a) and 5(b) show the in-plane resistance as a function of magnetic-field orientation for the standard and Corbino configurations, respectively. The in-plane resistance is expressed as sheet resistance, i.e., resistance per layer, normalized by $h/e^2$. We used $c = 1.75$ nm as the interlayer lattice constant. A remarkable difference in angular dependence is observed between the two geometries: in the standard configuration, the in-plane resistance exhibits a local minimum around θ ≈ 20°, whereas no such minimum is seen in the Corbino configuration. The most plausible interpretation of this difference is the presence or absence of edge-transport contributions in the $\nu = 0$ QHF state.

As seen in Fig. 5(a), the measured two-terminal in-plane sheet conductance in the standard geometry is much smaller than $2e^2/h$. This indicates that the edge transport is not ballistic but diffusive, due to frequent spin-inversion backscattering between the two helical edge channels. Such backscattering effectively divides the edge channel into many segments.

This behavior is similar to the dissipative edge transport caused by scattering at voltage contacts in the $v=0$ QH state of graphene [2]. The spin relaxation length (segment size) can be roughly estimated to be a few micrometers.

At the angle corresponding to the local minimum in Fig. 5(a), the in-plane MR exhibits strong saturation similar to that of the interlayer MR shown in Fig. 3. This suggests that the local minimum in the in-plane MR arises from mixing of resonant interlayer tunneling processes during edge scattering. The interlayer tunneling provides alternative conduction paths that bypass scattering sites.

In contrast, in the Corbino geometry there is no edge channel connecting the inner and outer electrodes. This is consistent with the absence of a local minimum in the angle dependence of the in-plane MR. In this geometry, the in-plane MR originates solely from the insulating bulk contribution, and indeed its temperature dependence exhibits insulating (thermally activated) behavior.

The clear difference in in-plane MR between the standard and Corbino geometries therefore provides strong evidence for the presence of edge transport in this system.

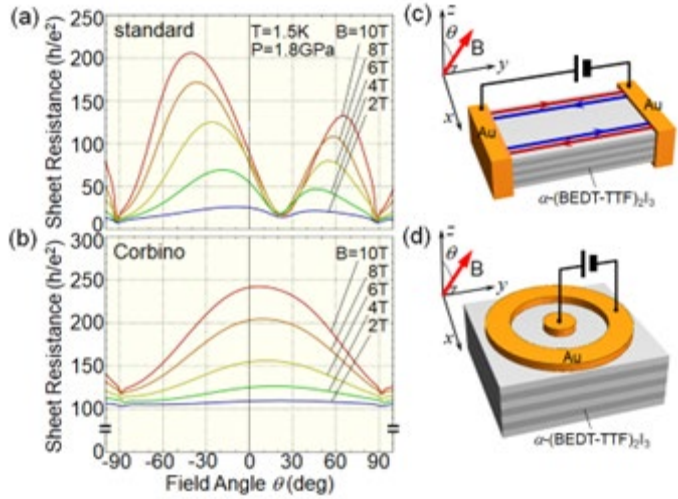

**Fig. 5.** In-plane two-terminal MR measured as a function of magnetic field orientation in the (a)(c) standard and (b)(d) Corbino geometries.

## 3. Discussion: possibility of the chiral magnetic effect

As an alternative mechanism for the local minimum observed in the angle-dependent MR, the chiral magnetic effect has been considered. Recent experiments suggest that $\alpha$-(ET)$_2$I$_3$ undergoes a dimensional crossover from a 2D Dirac-fermion system to a three-dimensional (3D) Dirac or Weyl semimetal at low temperatures, where interlayer coupling becomes non-negligible [7]. However, the nature of this 3D state has not yet been conclusively identified [8]. If the low-temperature 3D state is a type-II Weyl semimetal, as predicted theoretically [9], the chiral magnetic effect, which is specific to Dirac or Weyl semimetals, would appear only within a narrow range of magnetic-field orientations around the stacking axis [10]. In this limited angular range, the chiral magnetic effect would reduce the bulk MR (negative longitudinal MR), producing a dip-like structure in the angle dependence.

However, this possibility is ruled out by the present experimental results, particularly by the comparison between the standard and Corbino in-plane MR measurements, which clearly demonstrates that the dip structure originates from edge-channel transport rather than from a bulk mechanism such as the chiral magnetic effect.

## 4. Conclusions

In summary, our experiments provide strong evidence for the realization of the multilayer QHF state with helical edge states in $\alpha$-$(ET)_2I_3$. First, the high-field saturation of the interlayer MR does not scale with the sample cross-sectional area, indicating surface transport. Second, this saturation appears resonantly when the magnetic field is parallel to the side surface, consistent with resonant tunneling between edge states on adjacent layers. Third, the saturation persists up to 31 T, implying the robustness of the QHF state in the quantum limit. Finally, the in-plane MR shows a similar angle-dependent dip only in the standard geometry and not in the Corbino geometry, demonstrating that in-plane transport is also dominated by helical edge channels at the quantum limit.

## Acknowledgments

This work was partially supported by JSPS KAKENHI Grant Numbers JP23K03297 and JP24H01610. The author thanks Professor Tatsuo Hasegawa and Professor Yasuhiro H. Matsuda for providing an excellent research environment.